\documentclass{aiaa-pretty}

\author[Mishra and Mukhopadhaya and Iaccarino and Alonso]{ %
Aashwin Ananda Mishra\thanks{Research Fellow, Center for Turbulence Research, Stanford University, \texttt{aashwin@stanford.edu}},
Jayant Mukhopadhaya\thanks{Graduate Student, Department of Aeronautics \& Astronautics, Stanford University},
Gianluca Iaccarino\thanks{Professor, Center for Turbulence Research, Stanford University, AIAA Member.},
Juan Alonso\thanks{Professor, Department of Aeronautics \& Astronautics, Stanford University, AIAA Member.}\\
\textit{Stanford University, Stanford, California, USA}}

\title{An uncertainty estimation module for turbulence model predictions in SU2}

\begin{document}

\maketitle

\begin{abstract}
 .With the advent of improved computational resources, aerospace design has testing-based process to a simulation-driven procedure, wherein uncertainties in design and operating conditions are explicitly accounted for in the design under uncertainty methodology. A key source of such uncertainties in design are the closure models used to account for fluid turbulence. In spite of their importance, no reliable and extensively tested modules are available to estimate this epistemic uncertainty. In this article, we outline the EQUiPS uncertainty estimation module developed for the SU2 CFD suite that focuses on uncertainty due to turbulence models. The theoretical foundations underlying this uncertainty estimation and its computational implementation are detailed. Thence, the performance of this module is presented for a range of test cases, including both benchmark problems and flows relevant to aerospace design. Across the range of test cases, the uncertainty estimates of the module were able to account for a significant portion of the discrepancy between RANS predictions and high fidelity data.
\end{abstract}

\section{Introduction}

With the advent of ever more powerful computational resources over the last few decades, engineering design has moved from  predominantly a testing-based process to a simulation-driven methodology. One of the key ramifications of this transformation is a paradigm shift in dealing with uncertainties; those in operating conditions, input parameters, simulation models, etc. Traditionally, the quantification of uncertainties in engineering design was mainly based on the observed performance of similar systems over a period of time, theoretical studies, and various other ad hoc means and resulted in often unnecessary factors of safety and  conservativism. The design procedure has moved from exigencies such as factors of safety, built in redundancies, etc; in favor of explicit design under uncertainty to ensure consistent levels of reliability and performance. In light of these changes, the desired output from multi-physics simulations that support engineering design (such as computations of fluid flows, structural response, and heat transfer) are changing as well. In addition to improved fidelity from our simulations, we need explicit and reliable quantification of the uncertainties in our predictions and their associated margins of error

While these may arise from a multitude of sources, an important contributor to the errors and uncertainties in such simulations are the closure models used to estimate the effects of fluid turbulence.

Turbulent flows are found in numerous applications of aerospace interest where accurate prediction of flow parameters is an essential input to analysis and design. While computationally intensive approaches like Direct Numerical Simulations (DNS) and Large-Eddy Simulations (LES) offer high degrees of fidelity, their computational expense is prohibitive, preventing their adoption as a pragmatic tool in the engineering design process. Consequently, a vast majority of Computational Fluid Dynamics (CFD) studies of aerospace systems utilize simpler and relatively inexpensive Reynolds Averaged Navier-Stokes (RANS) closures to account for the effects of turbulence. 

Turbulence models are constitutive relations attempting to relate unknown quantities of interest (including higher-order statistical moments that may be expensive to compute exactly) to local, low-order flow statistics using simplifying assumptions. In this context, the goal of Reynolds-averaged Navier Stokes closures is to determine the Reynolds stress tensor as a function of mean flow quantities that are directly computable. To this end, RANS models utilize the concept of an isotropic eddy viscosity along with the modeling of turbulence processes via the gradient diffusion hypothesis. Due to their simplicity and cost-effectiveness, simple RANS models such as linear two-equation $k-\epsilon$ or $k-\omega$ closures are widely used in industrial applications. However, in spite of their wide spread use, RANS-based models suffer from an inherent structural inability to replicate fundamental turbulence processes. Due to the simplifications invoked in model formulation, RANS models can only represent certain features of turbulence with limited fidelity. For instance, in turbulent flows with significant effects of mean rotation, such as swirl or strong streamline curvature, the fidelity of linear eddy-viscosity-based closures is often unsatisfactory \cite{launder1987,craft1996}. In a congruous vein, the performance of two-equation models is found to be erroneous for cases with non-inertial frames of reference \cite{speziale1990}. In turbulent flows with flow separation and reattachment, eddy-viscosity-based models have had limited success. For instance, in turbulent flows in ducts, isotropic eddy-viscosity-based models are not able to reproduce the secondary flows that develop near the corners of the domain \cite{mompean1996}. Similarly, the eddy viscosity assumption is not valid in turbulent flows where turbulent time scales are significantly smaller than the mean shear time scale \cite{pope2001}.   

To aid in the establishment of RANS simulations as reliable tools for the engineering design process, there is need for Quantification of Margins and Uncertainties (QMU) in such models. In the most rudimentary manifestation, explicit quantification of the uncertainty in model predictions is needed in the form of prediction intervals for the Quantities of Interest.

While there has been an increase in the number of available CFD software packages over the years, at present, none of these offer an internal module for uncertainty estimation for the predictions. There are external packages that aid in estimating the effect of aleatoric uncertainties (those arising due to variabilities in input parameters and conditions), such as Dakota \cite{dakota}, NESSUS \cite{nessus}, etc. However, there are no reliable and extensively tested modules or libraries that address epistemic uncertainties, especially those arising due to the structural limitations of turbulence models. 

In this article, we outline the DARPA EQUiPS (Enabling Quantification of Uncertainty in Physics Simulations) module we have developed for the SU2 CFD suite \cite{su21}. The EQUiPS module focuses specifically on the estimation of uncertainties arising due to turbulence closure models.

We detail the theoretical foundations of the module and its computational framework. Thence, we validate the results of the module against high-fidelity data, across a set of test cases that include the entire spectrum from canonical benchmark flows to flows of relevance to aerospace design problems. 

During the development of this module, the overarching feature that we strive for is \textit{reliability}. This entails basing the computational module on rigorous theoretical foundations, along with extensive testing and validation to increase confidence in its results. Furthermore, a key feature of the module is its \textit{versatility}: its ability to cater to experts and non-experts alike. The end user for our computational tools may not be an expert at turbulence modeling or uncertainty quantification. For such a user, the manifestation of turbulent flow may be a small part of a much larger design problem. In this light, such a user needs a reliable, black-box tool that can provide interval estimates on the predictions of Quantities of Interest. EQUiPS is focused on such a user and the results outlined in this article correspond to the output that would be generated with the default settings of the module. Contrarily, a potential user with expertise in closure modeling may wish to customize the module to their specifications. The open-source character of the SU2 suite enables such an exercise. Another feature of the module is its \textit{computational pliability}. Contingent upon the computational capabilities of the user, the simulations of the module can be parallelized to reduce cumulative simulation time. Finally, a pragmatic parameter is the computational expense involved in obtaining reliable prediction intervals on the Quantities of Interest. If the requisite computational cost is extortionate, the choice of using a factor of safety or heuristic, empirical estimates to account for discrepancy in turbulence model predictions remains appealing. To inveigle new users to adopt this computational tool, we have ensured that this tool remains \textit{computationally inexpensive}. As is exhibited in the article, using the module, uncertainty estimates of engineering utility can be generated using just five perturbed RANS simulations.       

After an overview motivating this investigation in Section I, Section II outlines the mathematical and computational formalism. Herein, we explicate the Eigenspace Perturbation methodology for uncertainty estimation and detail the implementation in SU2. In Section III, the performance of this module is presented for a range of test cases, including both benchmark problems and flows relevant to aerospace design. The article concludes with a brief summary of results and future steps in Section IV.

\section{Computational and mathematical details}

\subsection{Eigenspace Perturbation Methodology}
The Eigenspace Perturbation methodology \cite{mishra} uses sequential perturbations on the eigenvalues and eigenvectors of the modeled Reynolds stress tensor to estimate turbulence model discrepancy. In the recent past, this framework has been successfully applied to a variety of canonical flows and benchmark cases \cite{emory2, mishra2016, xiaoghanem, xiao2, mishra5, gorleextra}. Utilizing the eigenvalue decomposition, the symmetric Reynolds stress tensor, $R_{ij}=\langle u_iu_j \rangle$, can be expressed as: $R_{ij}=2k (\frac{\delta_{ij}}{3}+v_{in}\Lambda_{nl}v_{lj})$, where $k$ denotes the turbulent kinetic energy, $v$ represents the eigenvector matrix, and, $\Lambda$, the diagonal matrix of eigenvalues of the Reynolds stress tensor. The tensors $v$ and $\Lambda$ are ordered such that $\lambda_1\geq\lambda_2\geq\lambda_3$. The amplitude, the shape and the orientation of the Reynolds stress are explicitly represented by $k$, $\lambda_i$ and $v_{ij}$, respectively. These perturbations are injected directly into the modeled Reynolds stress, expressed as:
\begin{equation}
R_{ij}^*=2k^* (\frac{\delta_{ij}}{3}+v^*_{in}\Lambda^*_{nl}v^*_{lj})
\end{equation}
where $^*$ represents the perturbed quantities. The perturbations to the eigenvalues, $\Lambda$, correspond to varying the componentiality of the flow (or the shape of the Reynolds stress ellipsoid). The projection of the eigenvalue perturbation in the barycentric map has both a direction and a magnitude. With respect to the direction of the eigenvalue perturbations, we focus on perturbations directed along the three vertices of the barycentric triangle: $x_{1C}, x_{2C}, x_{3C}$, each representing a limiting state of Reynolds stress anisotropy. The magnitude of the perturbation in the barycentric triangle is represented by $\Delta_{B} \in [0,1]$. Thus, $\Delta_{B}=0$ would leave the state unperturbed and $\Delta_{B}=1.0$, would perturb to the vertices of the Barycentric triangle. The perturbed barycentric coordinates, $\mathbf{x}^*$ , are given by: $\mathbf{x}^*=x+\Delta_B(\mathbf{x}^{(t)}-\mathbf{x})$, where $x^{(t)}$ denotes the target vertex (representing one of the one-, two- or three-component limiting states) and $x$ is the model prediction. For the computations in this investigation, instead of relying on a user-defined magnitude for $\Delta_B$, we set $\Delta_B=1.0$ so that the three limiting states are considered. These are the default settings of this parameter, $\Delta_B$, in the EQUiPS module and can be changed by the user. 

The perturbations to the eigenvectors, $v$, correspond to varying the alignment of the Reynolds stress ellipsoid. These are guided by the production of turbulent kinetic energy mechanism, $\mathcal{P}=-R_{ij}\frac{\partial U_i}{\partial x_j}$. The eigenvector perturbations seek to modulate turbulence production by varying the Frobenius inner product $\langle A,R \rangle =tr(AR)$, where $A$ is the mean gradient and $R$ is the Reynolds stress tensor. For the purposes of bounding all permissible dynamics, we seek the extremal values of this inner product. In the coordinate system defined by the eigenvectors of the rate of strain tensor, the critical alignments of the Reynolds stress eigenvectors are given by \cite{mishra} $v_{max}=\begin{bmatrix}
  1 & 0 & 0 \\
  0 & 1 & 0 \\
  0 & 0 & 1
 \end{bmatrix}$  and $v_{min}=\begin{bmatrix}
  0 & 0 & 1 \\
  0 & 1 & 0 \\
  1 & 0 & 0
 \end{bmatrix}$. The range of the inner products is $[\lambda_1\gamma_3+\lambda_2\gamma_2+\lambda_3\gamma_1, \lambda_1\gamma_1+\lambda_2\gamma_2+\lambda_3\gamma_3]$, where $\gamma_1 \geq \gamma_2 \geq \gamma_3$ are the eigenvalues of the symmetric component of $A$. 
 
\begin{figure}
\center
\includegraphics[width=0.75\textwidth]{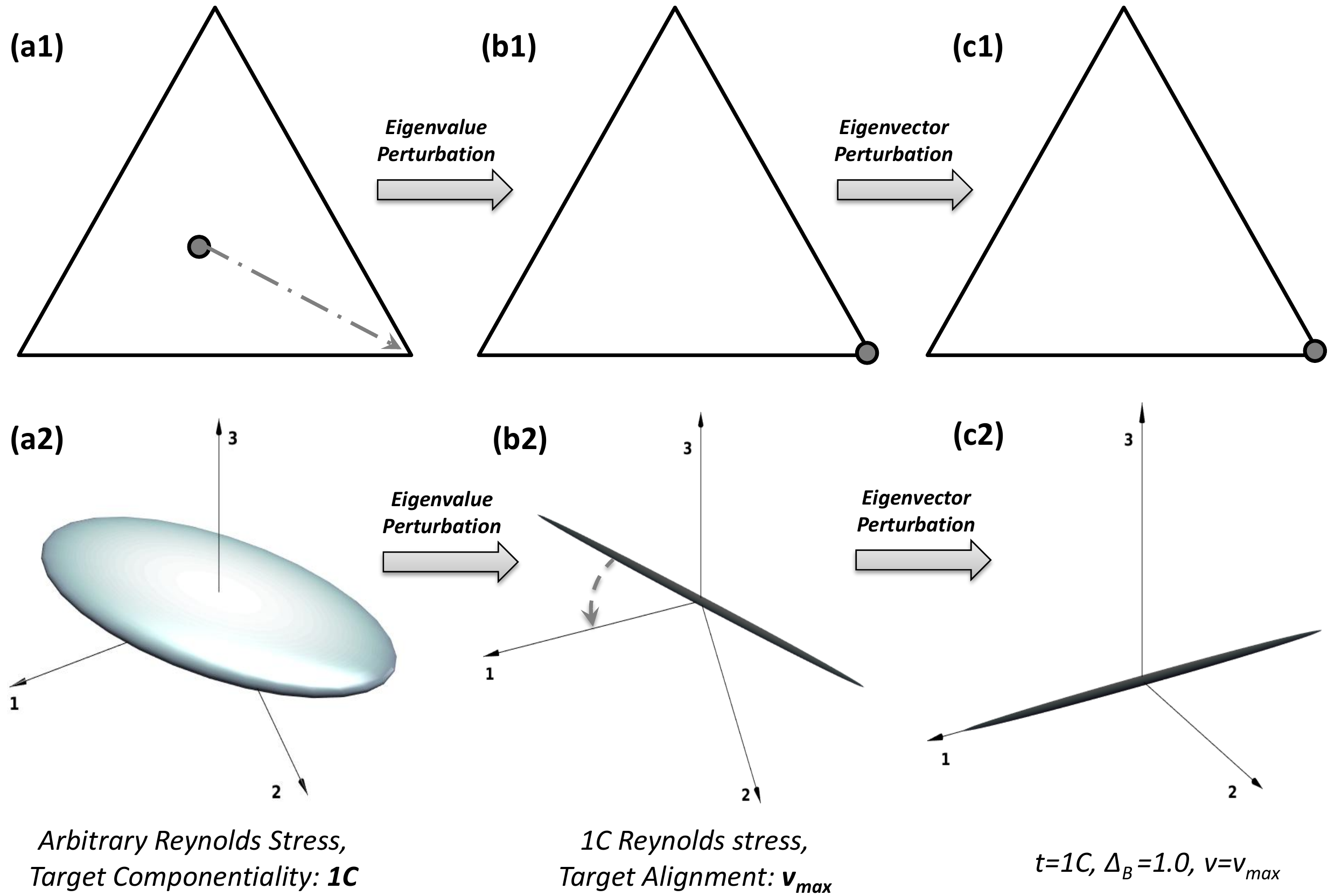}
\caption{Schematic outline of Eigenspace perturbations from an arbitrary state of the Reynolds stress. \label{fig:exp1}}
\end{figure}

In physical terms, the eigenvalue perturbation changes the shape of the Reynolds stress ellipsoid and the eigenvector perturbation changes its relative alignment with the principal axes of the mean rate of strain tensor. To illustrate this eigenspace perturbation framework, we outline a representative case schematically in figure \ref{fig:exp1}. In the upper row, $(a_1,b_1,c_1)$, we represent the Reynolds stress tensor at a specific physical location in barycentric coordinates and in the lower row, $(a_2,b_2,c_2)$, we visualize the Reynolds stress ellipsoid in a coordinate system defined by the mean rate of strain eigenvectors. These are arranged so that $\lambda_1 \geq \lambda_2 \geq \lambda_3,$ so essentially, the 1-axis is the stretching eigendirection and the 3-axis is the compressive eigendirection of the mean gradient.
 
Initially, the Reynolds stress predicted by an arbitrary model is exhibited in the first column, figure \ref{fig:exp1}, $a1$ and $a2$. The eigenvalue perturbation methodology seeks to sample from the extremal states of the possible Reynolds stress componentiality. Thus, we may, for instance, translate the Reynolds stress from this state to the $1C$ state, exhibited in the transition from figure \ref{fig:exp1}, $a1$ to $b1$. This translation changes the shape of the Reynolds stress ellipsoid from a a tri-axial ellipsoid to a prolate ellipsoid, exhibited in the transition from figure \ref{fig:exp1}, $a2$ to $b2$.
 
Thence, the eigenvector perturbation varies the alignment of this ellipsoid. Thus, we may, for instance, rotate the Reynolds stress ellipsoid so that its semi-major axis is aligned with the stretching eigendirection of the mean rate of strain tensor, exhibited in the transition from figure \ref{fig:exp1}, $b2$ to $c2$. This particular alignment would enable us to analyze impact of the maximum permissible production on turbulence evolution. In conjunction, these two perturbation approaches enable us to maximize the information we may get from single-point statistics to quantify uncertainty bounds.

\begin{figure}
\center
\includegraphics[width=0.8\textwidth]{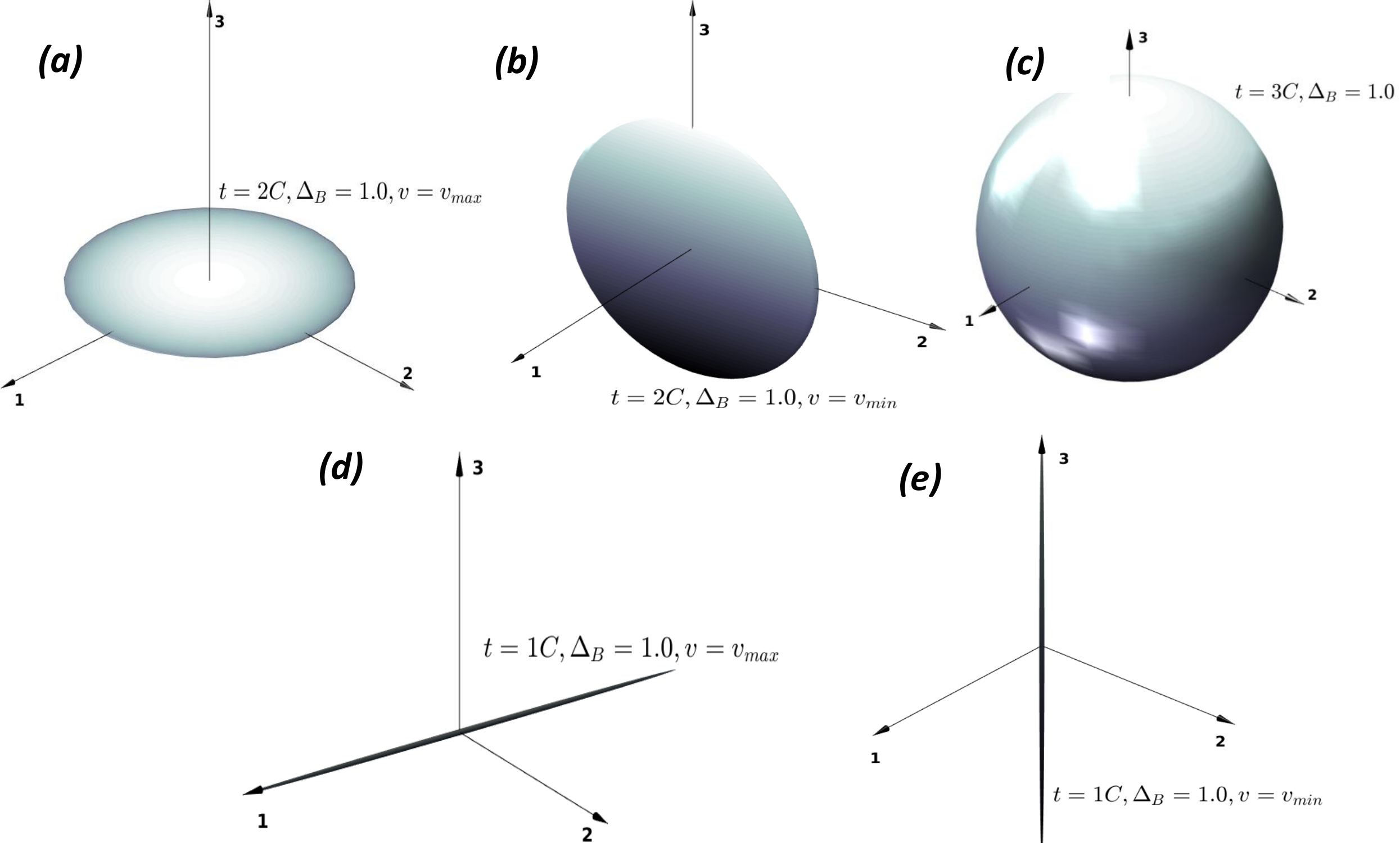}
\caption{Schematic visualization of the extremal states, as Reynolds stress ellipsoids, in the eigenspace perturbation methodology.\label{fig:exp2}}
\end{figure}

This eigenspace perturbation framework gives us $5$ distinct extremal states of the Reynolds stress tensor, these are schematically displayed in figure \ref{fig:exp2}. These correspond to 3 extremal states of the componentiality $(1C, 2C, 3C)$ and 2 extremal alignments of the Reynolds stress eigenvectors, $(v_{min}, v_{max})$.  For the $3C$ limiting state, the Reynolds stress ellipsoid is spherical. Due to rotational symmetry, all alignments of this spherical Reynolds stress ellipsoid are identical and eigenvector perturbations are superfluous. Thus, for uncertainty bounds on flow evolution engendered by eigenspace perturbations, we need a set of 5 RANS simulations. 

In the succeeding section, we contrast the baseline predictions of the RANS closure along with uncertainty bounds against high fidelity data. The uncertainty bounds on the profiles are engendered by the union of all the states lying in the set of perturbed RANS simulations. This set of five different eigenspace perturbations can be designated as $ \left\{ (1C,v_{max}), (1C,v_{min}), (2C,v_{max}), (2C,v_{min}), (3C,v_{max}/v_{min}) \right\}$. Together, these perturbations account for the extreme cases of Reynolds stress componentiality and eigenvector alignments. Due to the nature of this formulation, these bounds are explicitly deterministic.

\subsection{The Stanford University Unstructured CFD suite}

The SU2 software suite was developed with an emphasis on Partial Differential Equation (PDE) analysis along with PDE-constrained optimization problems while utilizing unstructured meshes \cite{su21}. Although the framework is general and meant to be extensible to any sets of governing equations for multi-physics problems, its focus is a Reynolds Averaged Navier-Stokes (RANS) solver capable of simulating compressible, turbulent flows found in aerospace engineering and design problems. To ensure the reliability of the baseline SU2 RANS solver, extensive validation studies of the SU2 platform have been conducted across a diverse and diametric assortment of turbulent flows \cite{su22}. 

SU2 is being actively developed in the Aerospace Design Lab of the Department of Aeronautics and Astronautics at Stanford University. It has been released under an open-source license and is freely available to the community, so that developers may contribute to the source code and further improve the accuracy and capabilities of the suite. The modular structure of SU2, wherein modules share common sets of classes and data structures within an object-oriented code architecture, makes it an ideal vehicle for multi-physics simulations and aerodynamic shape optimization. To accomplish this, the SU2 development team has included industry-standard solver technology for turbulent flows while also developing numerical solution algorithms that result in robust rates of convergence. Finally, SU2 includes continuous adjoint solver implementations for efficiently computing shape design gradients that can be further improved utilizing contributions from the developer community.  

\subsection{EQUiPS: Enabling Quantification of Uncertainty in Physics Simulations}
\begin{figure}
\centering
\includegraphics[width=1.05\textwidth]{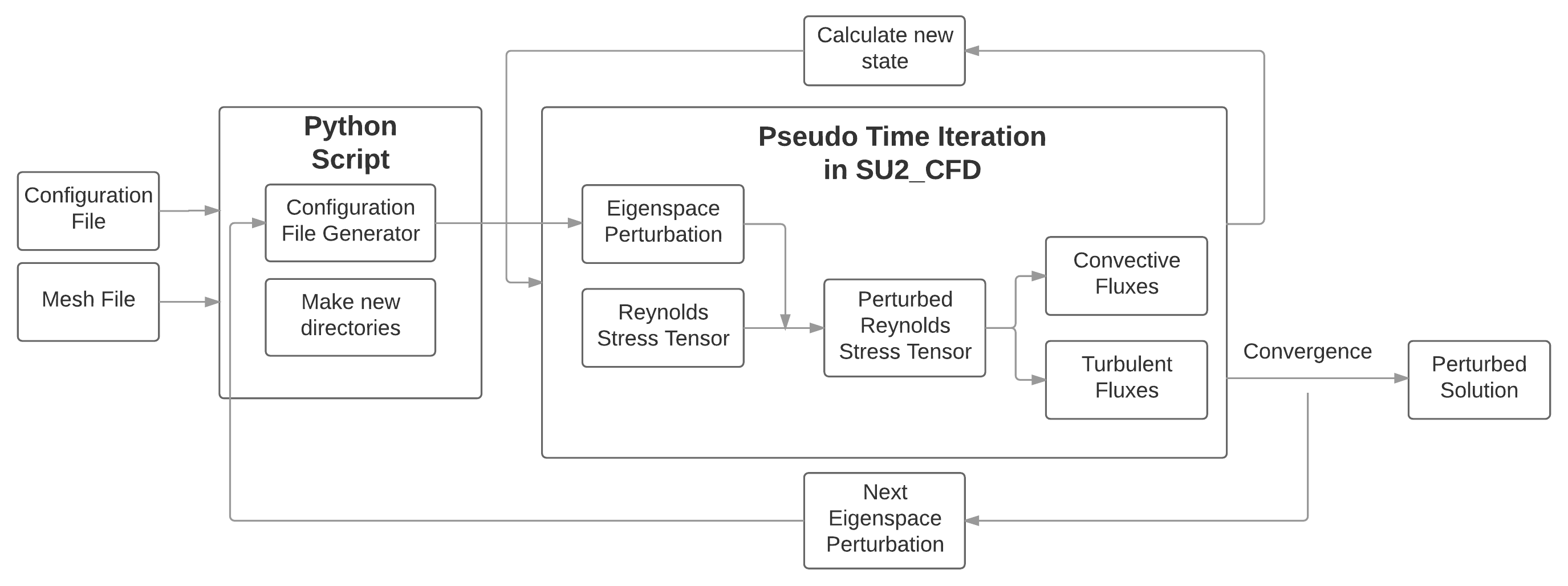}
\caption{Flow chart showing the implementation of EQUiPS within SU2 \label{fig:flowChart}}
\end{figure}
SU2's architecture lends itself to easy integration of turbulence model uncertainty estimation framework without significant alterations to the code. The module is broken up into two parts: a high-level python script that sequentially specifies the perturbation to be performed, and C++ code that is added to the existing code base that performs the perturbations during the execution of simulations.

For smooth operation, it is best to have performed a baseline simulation with SU2 and have achieved sufficient convergence. This ensures that the mesh file, and the input configuration file are well posed, and, if run through the python script, can provide converged perturbed solutions. A high level representation of the module and how it fits within the SU2 architecture is shown Fig. \ref{fig:flowChart}. 

The python script takes an input configuration file, and a mesh file that are identical to ones that would be used to run the baseline CFD simulation in SU2. The script sets the necessary configuration options to run the EQUiPS framework. These include the direction of the eigenvalue perturbation (aligned towards one of the $1C, 2C$ or $3C$ state on the barycentric triangle for a specific simulation), the magnitude of the eigenvalue perturbation ($\Delta_{B} \in [0,1]$) and the perturbed alignment of the Reynolds stress eigenvector ($v_{min}$ or $v_{max}$, as detailed earlier in this section).  It sequentially runs through the simulations corresponding to perturbed states, creating a new directory for each new simulation, and outputting the results in the respective directories. These results, in conjunction with the previously run baseline solution, can be post-processed to provide the necessary model form uncertainty information.

The implementation within the existing code base is limited to the two areas where the perturbations are injected into the simulation: in the viscous flux calculations, and in the turbulent flux calculations. For each node in the mesh, the baseline Reynolds stress tensor is calculated and decomposed into its eigenvalues and eigenvectors. Using the eigenspace perturbation methodology shown earlier, the eigenvalues are perturbed as prescribed, and, if needed, the eigenvectors are permuted. The same eigenspace perturbation is used for all the nodes in the domain. An under-relaxation factor is used to gradually march the stress tensor to the perturbed state. This makes the simulation numerically stabler.

The new eigenvalues and eigenvectors are then recomposed into a new stress tensor for each node. These perturbed stress tensors are then used to calculate the viscous fluxes to advance each node to the next pseudo-timestep. A similar procedure of decomposition, perturbation, and re-composition is followed in the calculation of the turbulent fluxes. In the case of the SST turbulence model \cite{sst}, the effects of the eigenvector perturbation manifests itself in the turbulence production term. 

The new perturbed fluxes are then used to march the solution forward in pseudo time. As the solution converges, the Reynolds Stress also converges to its perturbed state. Once a perturbed solution is converged, the python script moves on to the next eigenspace perturbation to be performed. It creates a new sub-directory and a new configuration file to specify the perturbation to be performed. This continues until all the specified perturbed simulations are complete. 

In the spirit of \textit{versatility}, the implementation was developed with the aim of reducing the number of user-defined inputs as much as possible. This allows the framework to be used without in-depth knowledge of the mechanics behind it. The details of the perturbations (componentality, eigenvector permutation etc.) are abstracted away to provide a clean user interface that does not deviate from the work flow of running a regular CFD simulation. For the potential expert user, all the options used for the module can also be specified in the configuration file, without the need for the python script. 

\section{Results and discussion}

In this section, we outline results for the uncertainty estimation module for a range of test cases detailed in Table 1. The first two flows are benchmark cases widely used for the testing of turbulence models. The final four cases are flows of interest to aerospace design. For the results, the perturbation parameters were retained at their default values, so the bounds correspond to the set $ \left\{ (1C,v_{max}), (1C,v_{min}), (2C,v_{max}), (2C,v_{min}), (3C,v_{max}/v_{min}) \right\}$. The SST model of \cite{sst} was used for all the results. 

\begin{table}
\caption{\label{tab:table1} Details of testing and validation cases}
\begin{center}
\begin{tabular}{ccc}
Test Case& Rationale& Notes \\\hline
Turbulent flow over a backward facing step& Benchmark flow& 2D, incompressible simulation.\\
Turbulent flow through an asymmetric diffuser& Benchmark flow& 2D, compressible simulation.\\
Jet effluxes from the NASA Acoustic Response Nozzle& Engineering case & 3D, sub-sonic simulation.\\
Cooled Jet via the NASA Seiner Nozzle& Engineering case& 3D, super-sonic simulation.\\
NACA 0012 airfoil at a range of angles of attack& Engineering case& 2D, sub-sonic simulation.\\
Turbulent flow over a three-element High-lift airfoil& Engineering case& 2D, sub-sonic simulation.\\
\end{tabular}
\end{center}
\end{table}

\subsection{Turbulent flow over a backward facing step}
\begin{figure}
\centering
\includegraphics[width=0.8\textwidth]{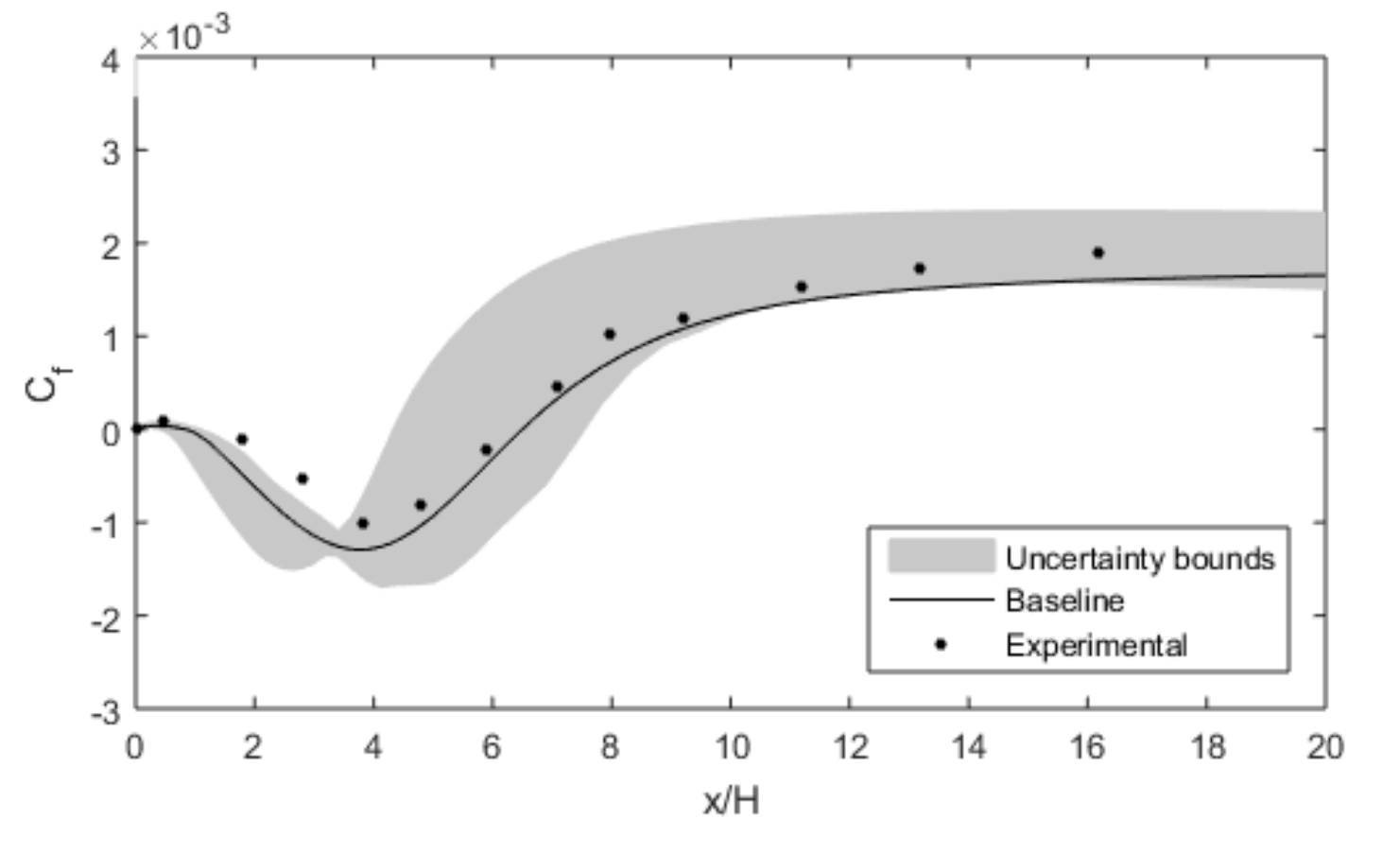}
\caption{Coefficient of friction ($c_f$) over the bottom wall for the flow over a backward facing step. The filled circles mark the experimental data of \cite{driver1985}, the solid line represents the unperturbed RANS simulation and the gray shaded zones are the uncertainty bounds\label{fig:fig1}}
\end{figure}

\begin{figure}
\centering
\includegraphics[width=0.8\textwidth]{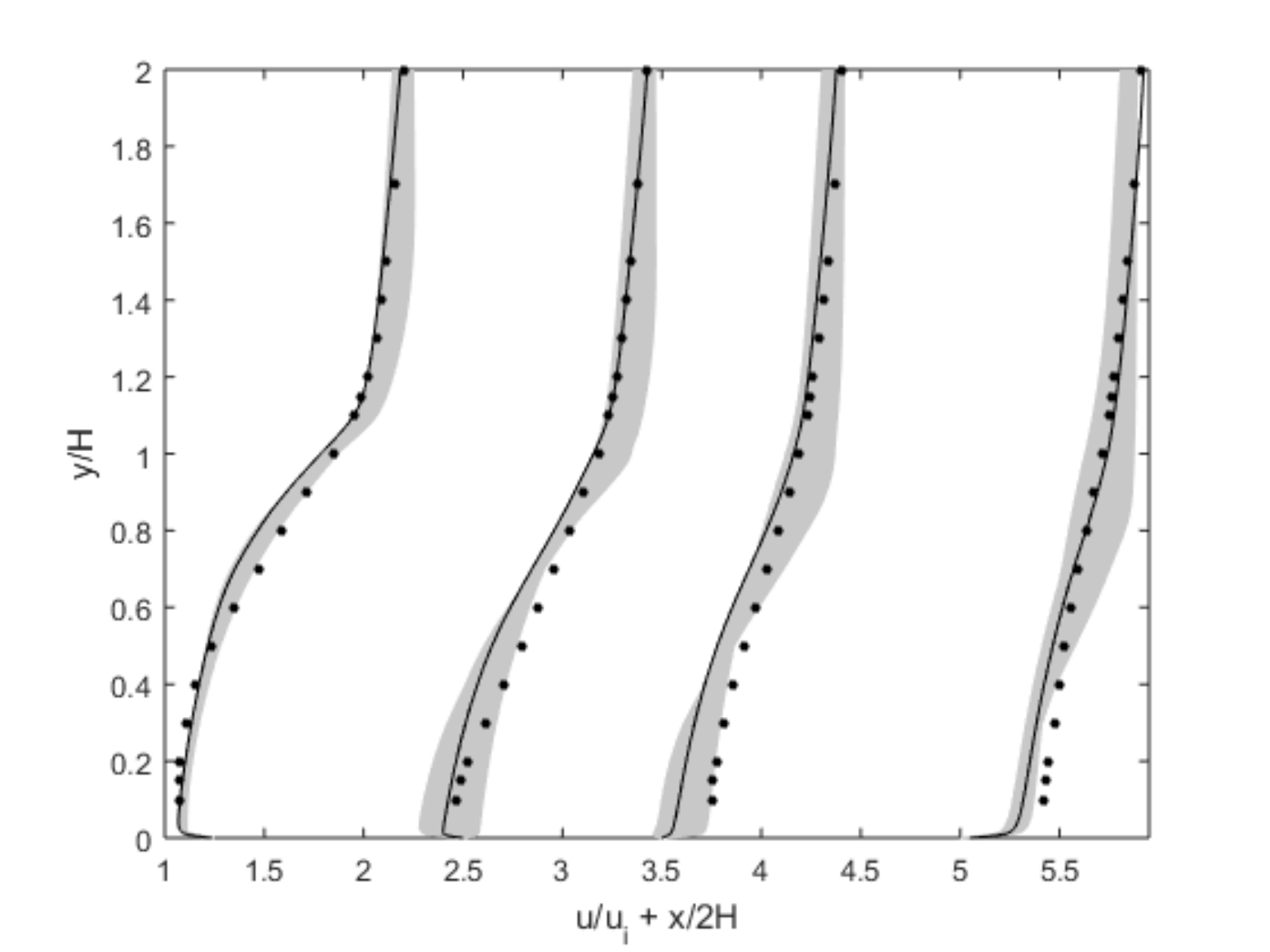}
\caption{Mean velocity profiles for the flow over a backward facing step. The filled circles mark the experimental data of \cite{driver1985}, the solid line represents the unperturbed RANS simulation and the gray shaded zones are the uncertainty bounds\label{fig:fig2}}
\end{figure}

Turbulent flow over a backward facing step involves flow separation, followed by the reattachment of separated shear layers under the influence of adverse pressure gradients. The location of flow separation is fixed by the geometry and the point of reattachment is contingent upon the pressure gradient. The velocity gradients in the flow field cause turbulent production far off the wall region and their interaction with the mean flow influences the size of the separation bubble after the step. The extent of this bubble or the re-attachment length is a key quantity that must be predicted accurately by a turbulence model. Classical two-equation turbulence models under predicted the re-attachment length by a substantial amount of the order of over 10-25\% \cite{thangam1991}.This occurs due to inaccurate predictions for normal Reynolds stress differences arising from the use of an isotropic eddy viscosity \cite{thangam1991}. This configuration has been investigated experimentally by \cite{driver1985} and their data is used in our investigation.

Fig. \ref{fig:fig1} outlines coefficient of friction ($c_f$) along the bottom wall. The uncertainty estimates are able to account for the discrepancy between model predictions and DNS data at most locations. Fig. \ref{fig:fig2} outlines the axial mean velocity profiles in the flow, after the step. While there is some discrepancy between the unperturbed RANS simulation and the experimental data, the uncertainty intervals are able to account for most of this discrepancy with almost all experimental data points lying in the uncertainty bounds. 

\subsection{Turbulent flow through an asymmetric diffuser}

\begin{figure}
\centering
\subfigure[\label{fig:fig3a}Coefficient of friction ($c_f$) over the top wall ] %
  {\includegraphics[width=0.8\textwidth]{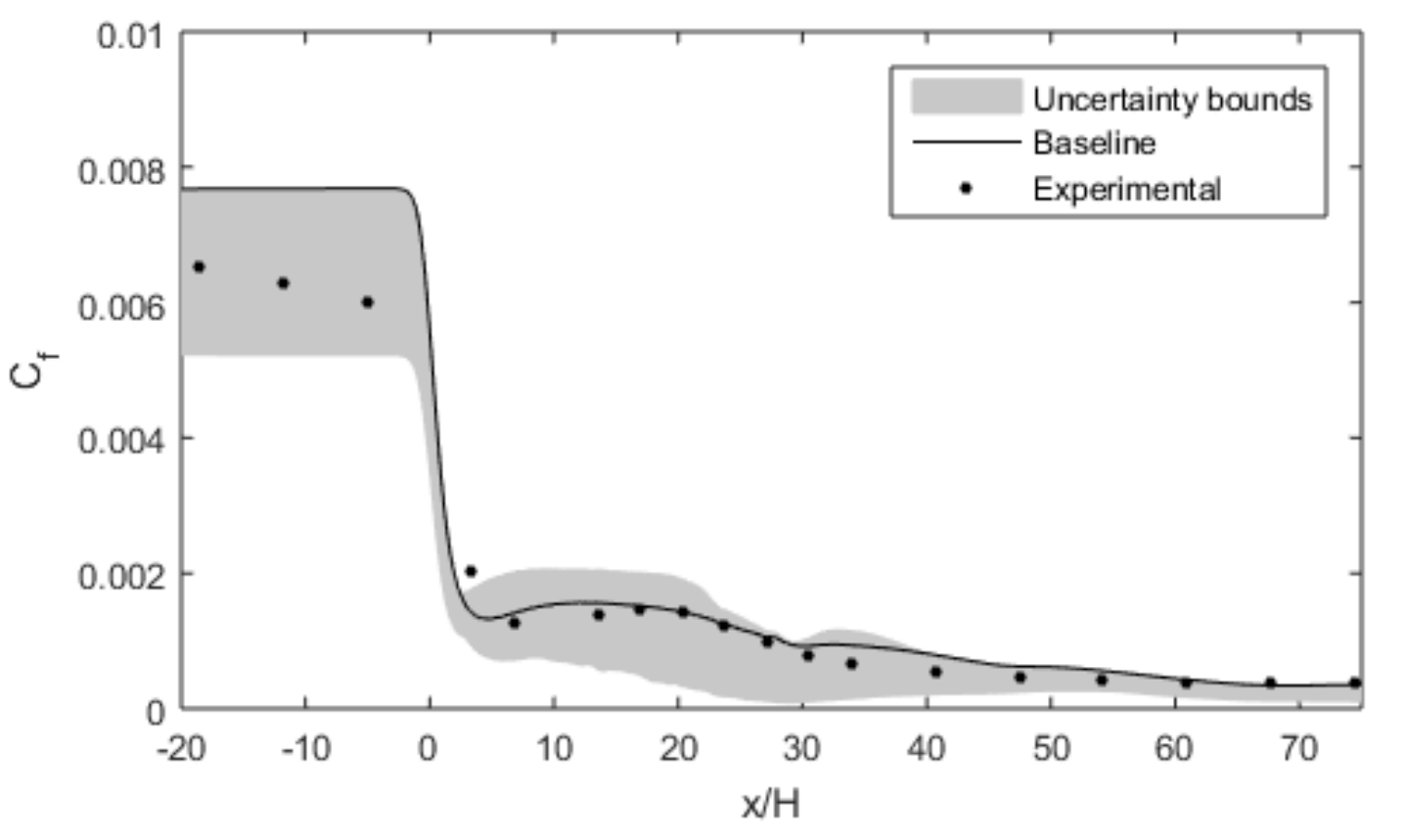}}
\subfigure[\label{fig:fig3b}Coefficient of friction ($c_f$) over the bottom wall ] %
  {\includegraphics[width=0.8\textwidth]{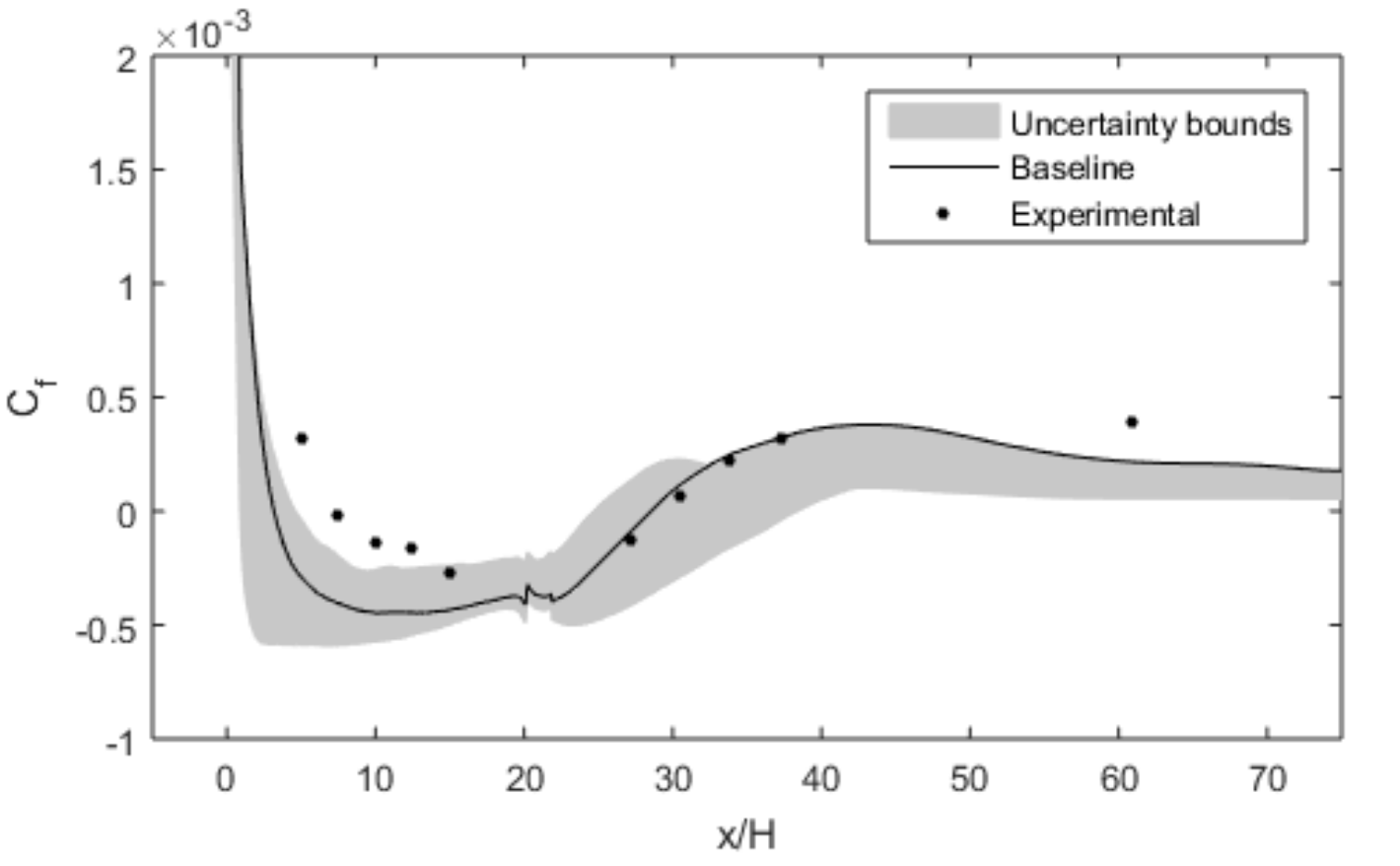}}
\caption{Coefficient of friction ($c_f$) over the top and bottom walls for the flow in an axisymmetric diffuser. The filled circles mark the experimental data of \cite{buice}, the solid line represents the unperturbed RANS simulation and the gray shaded zones are the uncertainty bounds}
\end{figure}

Turbulent flow in an axisymmetric diffuser has many interesting properties, such as separation over a smooth wall, subsequent reattachment and redevelopment of the downstream boundary layer, that offer considerable challenges to eddy-viscosity based models. We use the experimental data of \cite{buice}. The data include mean velocities at various stations in the diffuser and skin friction data on the upper and lower walls. For the simulations, the inlet conditions are specified as a fully-developed channel flow at $Re=20,000$ based on the centerline velocity and the channel height. 

\begin{figure}
\centering
\includegraphics[width=\textwidth]{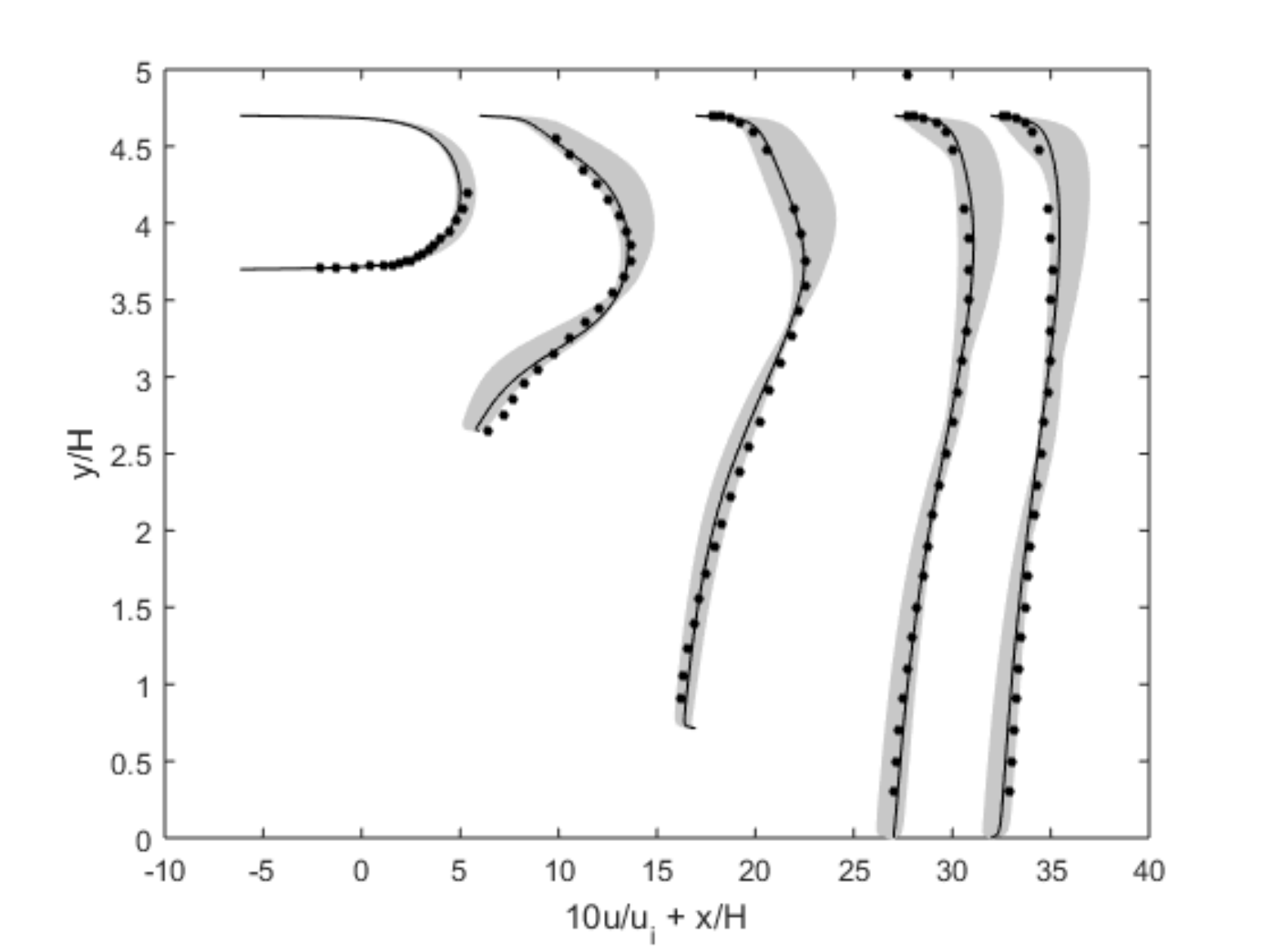}
\caption{Mean velocity profiles for the flow in the axisymmetric diffuser. The filled circles mark the experimental data of \cite{driver1985}, the solid line represents the unperturbed RANS simulation and the gray shaded zones are the uncertainty bounds\label{fig:fig4}}
\end{figure}

Fig. \ref{fig:fig3a} and \ref{fig:fig3b} exhibit the coefficient of friction ($c_f$) over the top and bottom walls. While there is considerable discrepancy between the RANS prediction and the experimental data, the uncertainty intervals are able to account for a substantial proportion of this discrepancy. Most of the experimental data points lie inside the uncertainty intervals. Fig. \ref{fig:fig4} exhibits the axial mean velocity profiles at a range of locations along the diffuser. The uncertainty bounds encompass most of the experimental data points, including in the region of flow separation. 

\subsection{Jet efflux of the NASA Acoustic Response Nozzle}

\begin{figure}
\includegraphics[width=1.1\textwidth]{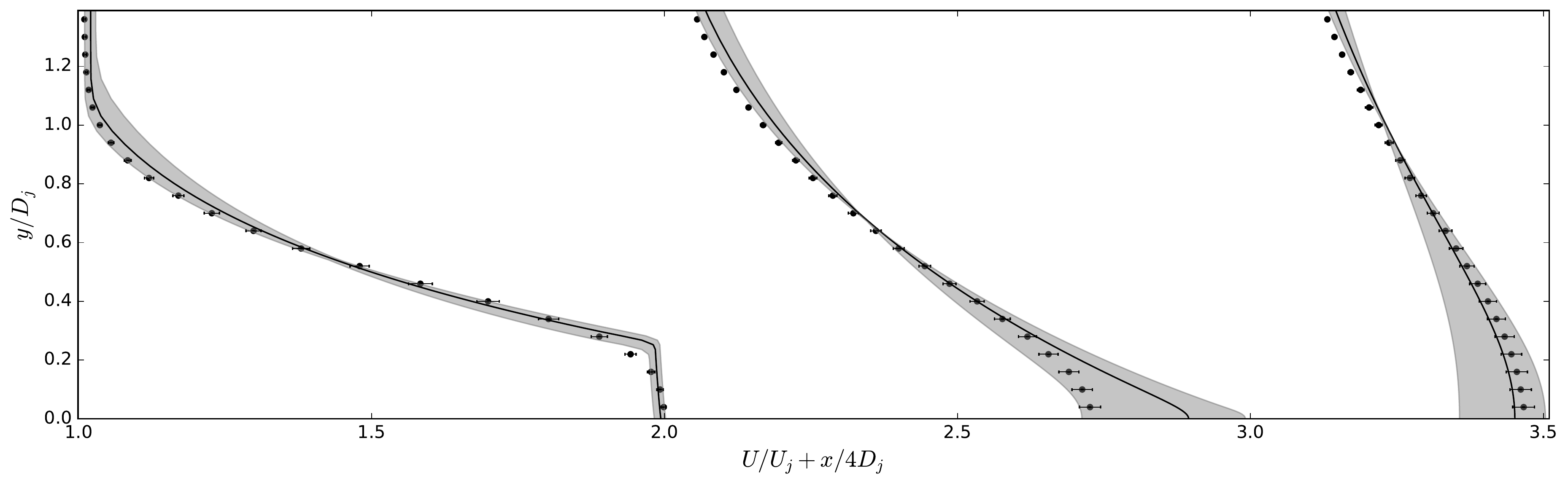}
\caption{Mean axial velocity profiles ($U$) in the NASA Acoustic Response Nozzle heated jet at $Ma=0.376$, delineated at $x/D_j= 4,8$ and $12$\label{fig:fig5a}}
\end{figure}

\begin{figure}
\includegraphics[width=1\textwidth]{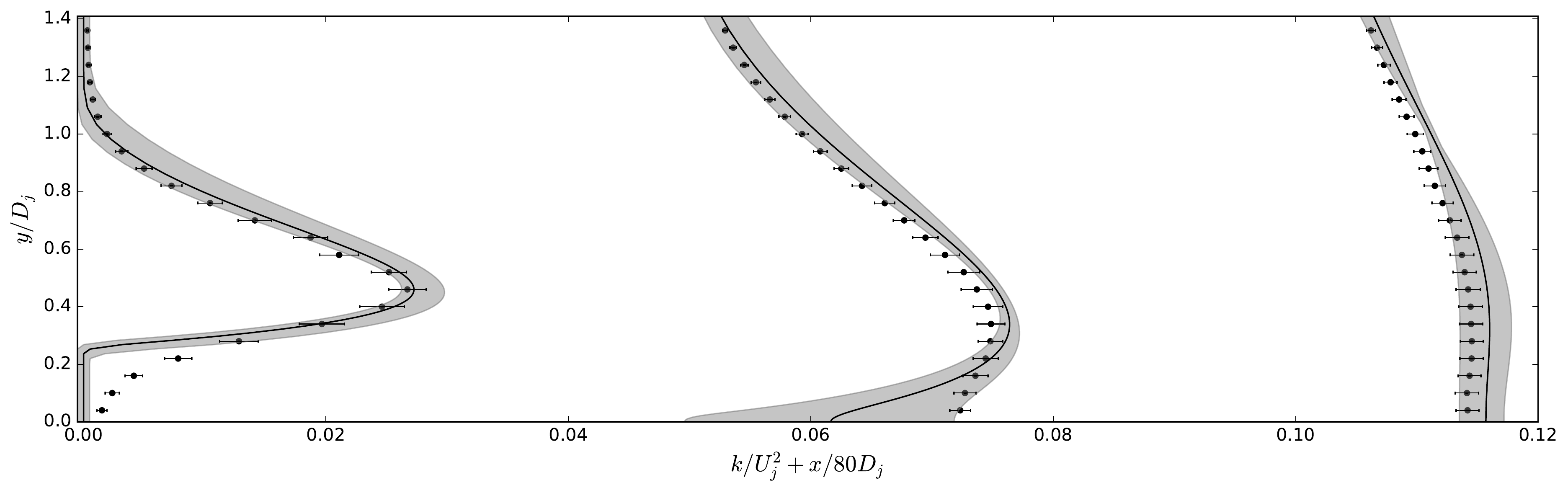}
\caption{Turbulence kinetic energy ($k$) profiles in the NASA Acoustic Response Nozzle heated jet at $Ma=0.376$, delineated at $x/D_j= 4,8$ and $12$\label{fig:fig5b}}
\end{figure}

Reliable predictions of turbulent jets exhausting from contoured aircraft nozzles are essential for a variety of aerospace design applications. However, these exhaust jets involve a multitude of complications such as the interaction of the jet efflux and the ambient, complicated nozzle geometries, compressibility effects, under or over-expanded flows, etc. These pose significant challenges to eddy-viscosity based models. As an illustration, we may focus on the mixing between the jet and the ambient fluid. In the vicinity of the nozzle exit, RANS models predict a significantly lower rate of jet mixing as compared with high-fidelity data. Farther downstream of the jet potential core, RANS models predict the far-field mixing rate to become substantially higher than is observed in experiments. Similarly, the fidelity of RANS predictions is highly inconsistent, having higher fidelity for cold jets than heated, for axisymmetric than nonaxisymmetric geometries, varying significantly over different quantities of interest, etc.

In this test case, we investigate sub-sonic jets from the NASA Acoustic Response Nozzle. This has been studied experimentally by \cite{nasajet} and extensive Particle Image Velocimetry data are available. Along with the sample mean of Quantities of Interest over repeated realizations of the experiment, the standard deviations are also available. While different jets with diverse flow parameters were tested, in the interest of brevity, we outline results for 2 cases, detailed in Table 2.

\begin{table}
\caption{\label{tab:table2} Reference conditions for subsonic jet flow cases considered}
\begin{center}
\begin{tabular}{cccc}
Test Case& Classification& $Ma$& ${T_0}/{T_{0,\infty}}$\\\hline
Case I& heated, perfectly expanded& 0.376& 1.764\\
Case II& cooled, perfectly expanded& 0.513& 0.950\\
\end{tabular}
\end{center}
\end{table}

\begin{figure}
\includegraphics[width=1.05\textwidth]{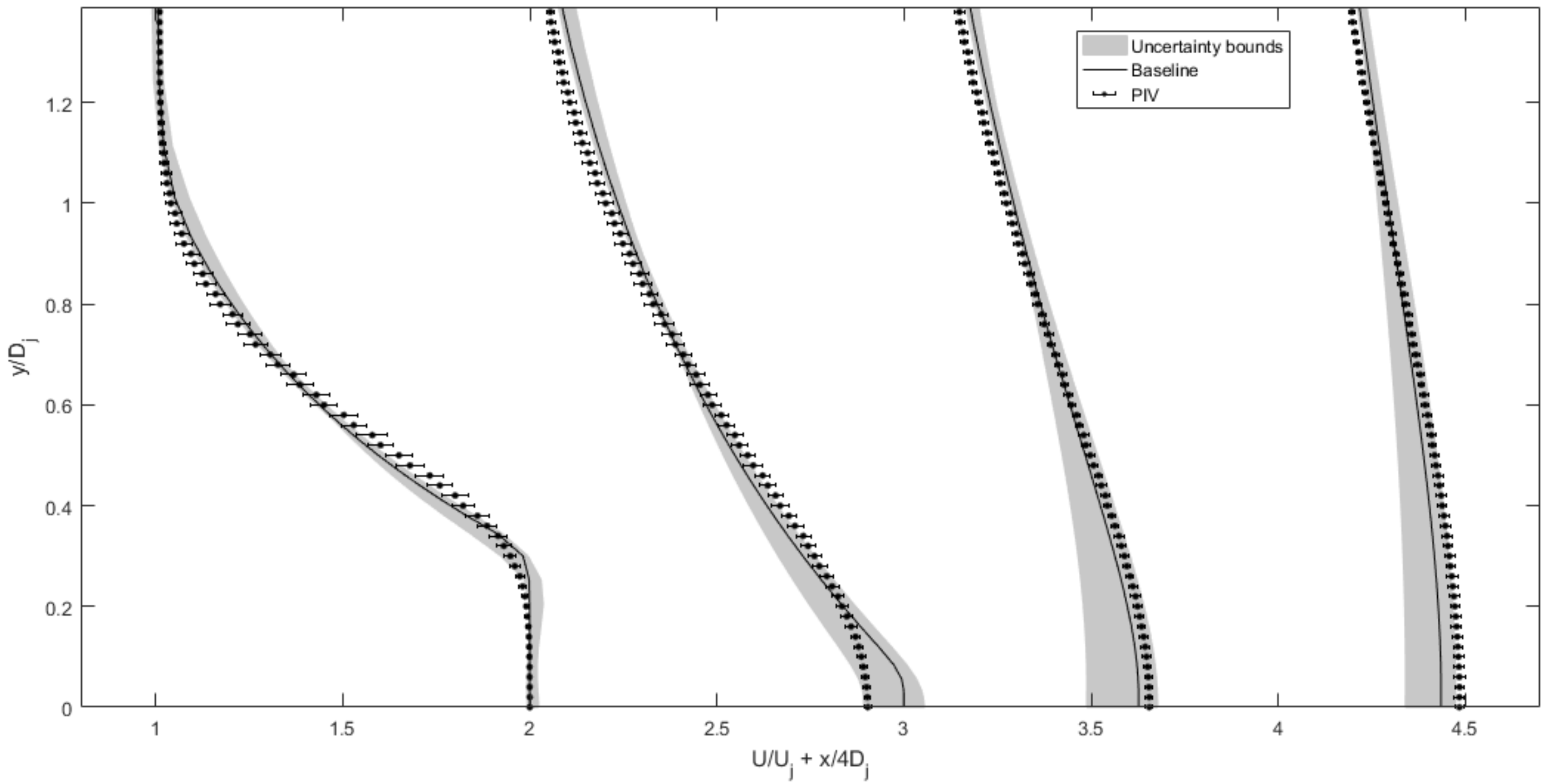}
\caption{Mean axial velocity profiles in the NASA ARN exhaust , delineated at $x/D_j= 4,8,12$ and $16$\label{fig:fig5}}
\end{figure}

The mean axial velocity profiles at $x/D_j= 4,8$ and $12$ are outlined in Fig. \ref{fig:fig5a} for Case I from Table 2. The error bars associated with the experimental data represent the standard deviation over multiple realizations of the experiment. The uncertainty bounds are able to account for a significant proportion of the discrepancy between the RANS simulations and the high fidelity data. Almost all the experimental data points are contained in the uncertainty bounds. 

Fig. \ref{fig:fig5b} exhibits the turbulent kinetic energy profiles at $x/D_j= 4,8$ and $12$ for Case I from Table 2. The uncertainty bounds are able to account for a significant ratio of the difference between the RANS predictions and the PIV measurements. At most locations, the uncertainty bounds envelope the experimental data, or have a non-trivial intersection with the associated error bars. A similar trend is observed in Fig. \ref{fig:fig5} for Case II from Table 2.

\subsection{Heated Jet through the NASA Seiner Nozzle}

\begin{figure}
\subfigure[\label{fig:fig7a} Mach number variation over the centerline] %
  {\includegraphics[width=0.5\textwidth]{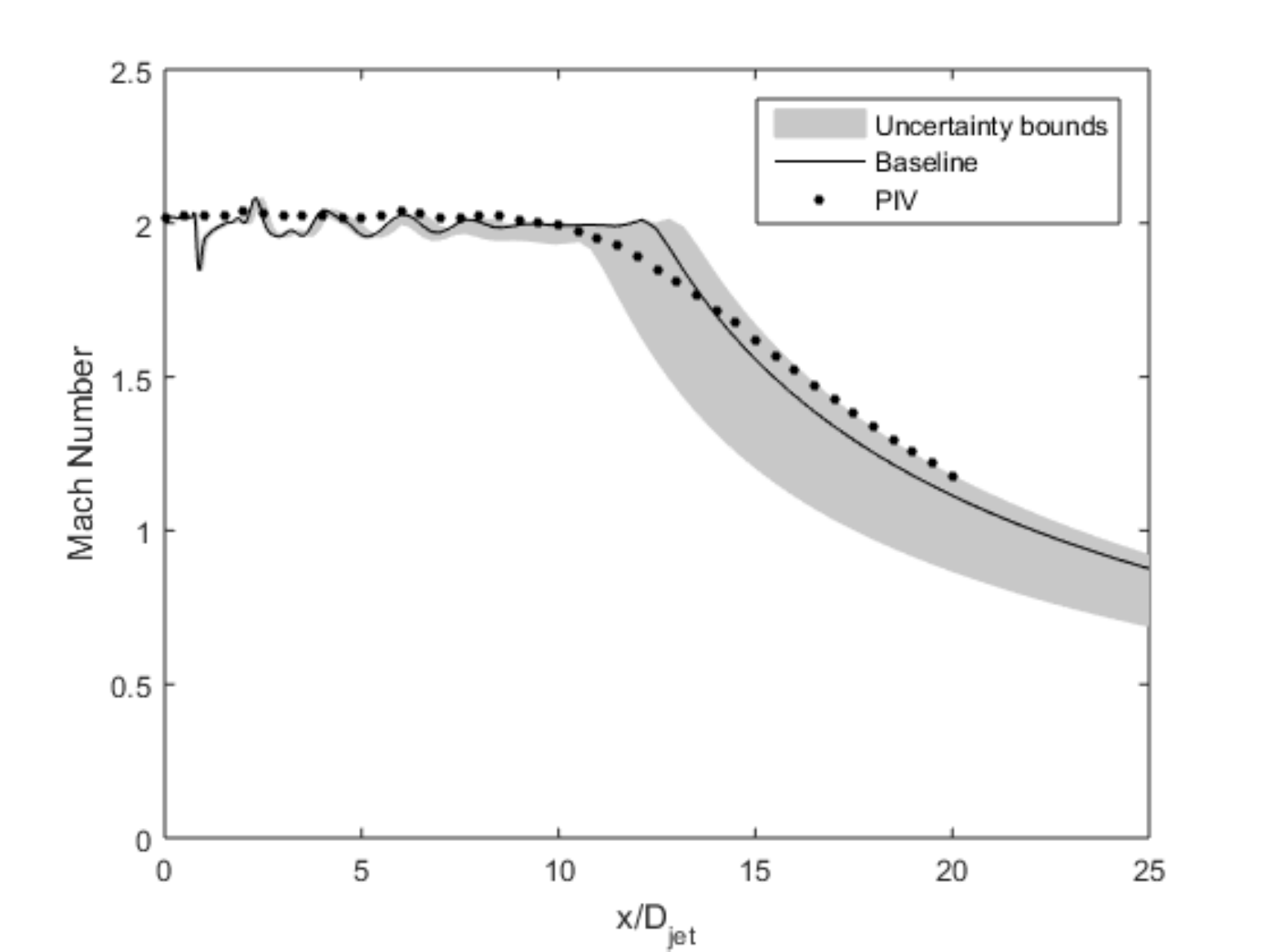}}
\subfigure[\label{fig:fig7b}Pressure variation over the centerline] %
  {\includegraphics[width=0.5\textwidth]{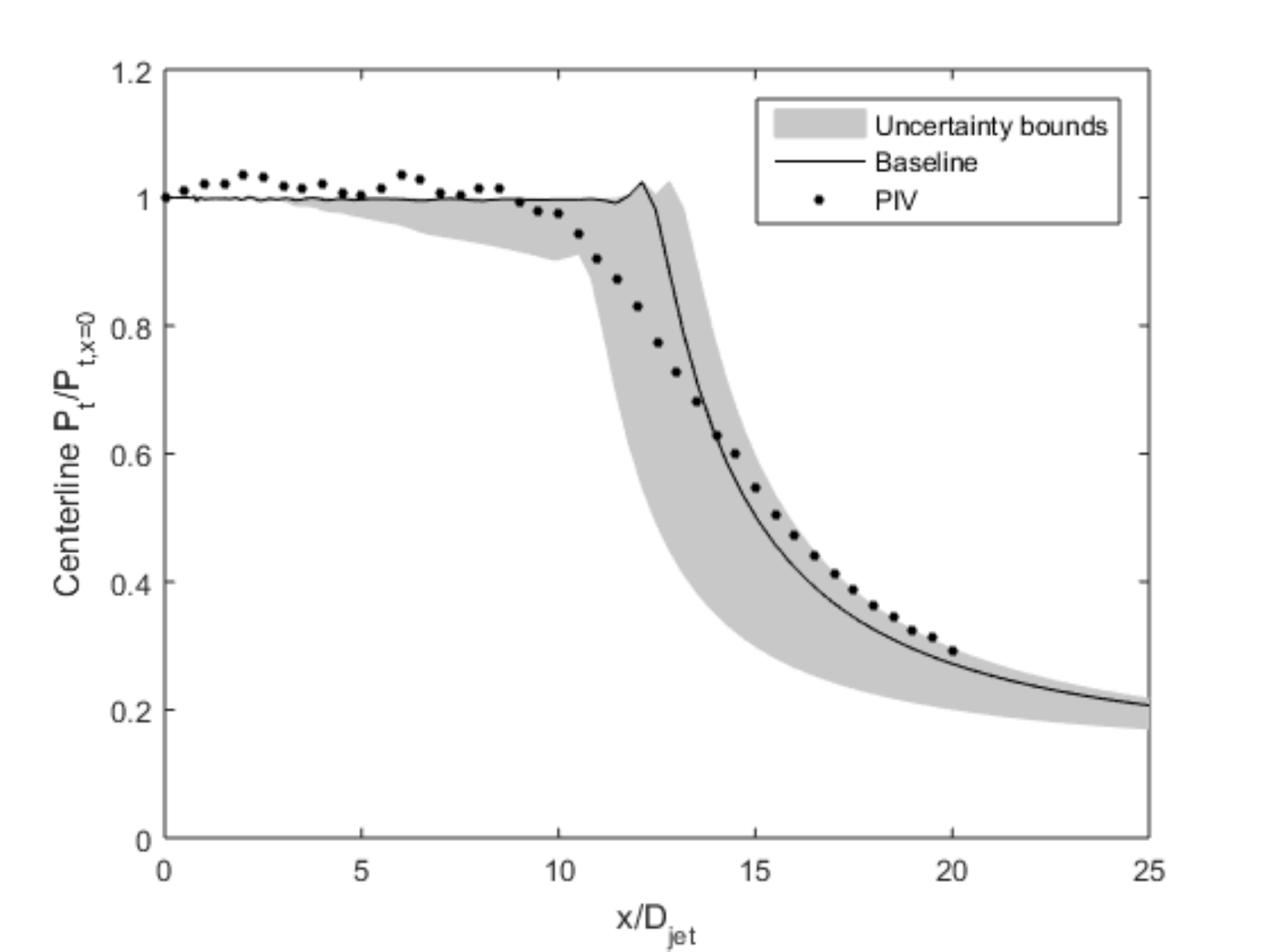}}
\caption{Variation in Quantities of Interest over the jet centerline for the cooled supersonic jet \cite{seiner}.}
\end{figure}

In this test case, we transition from sub-sonic to super-sonic jets. Herein, we replicate the experiment configuration of \cite{seiner}. The experiment was conducted on an axisymmetric, convergent-divergent nozzle with diameter $9.144 cm$. The jet exhausted into a quiescent ambient, with  $Ma=2.0$ and Reynolds number $1.3 \times 10^6$. Fig. \ref{fig:fig7a} and \ref{fig:fig7b} outline the variation in the Mach number and pressure along the jet centerline. Fig. \ref{fig:fig7a} indicates that the extent of the jet potential core is over predicted by the RANS model. The turbulence model also over predicts the spreading rate of the jet. This indicates a predicted rate of mixing that is higher than is suggested by the experimental data. However, the uncertainty bounds are able to account for most of the discrepancy in the length of the potential core and the ensuing mixing rate, with almost all the experimental data points lying inside the bounds. 

\subsection{NACA 0012 airfoil at a range of angles of attack}
In this test case, we consider the flow over a NACA 0012 airfoil at a range of angles of attack, varying from 0$^{\circ}$ to 20$^{\circ}$. This design was chosen specifically due to its adoption in the industry. These include in conventional aircraft for instance the wing tips of the Cessna 140A, 207; helicopter designs such as the inboard and outboard blades of the Aerospatiale AS365, Boeing 600N, Lockheed 475; in addition to numerous horizontal and vertical axis wind turbines. The Reynolds number for the simulations is $6\times 10^6$. The flow is sub-sonic with $Ma=0.15$. The experimental data \cite{ladson1988} report the coefficient of lift ($C_L$) and the surface pressure coefficient ($C_P$) for different angles of attack. 

\begin{figure}
\center
\includegraphics[width=0.7\textwidth]{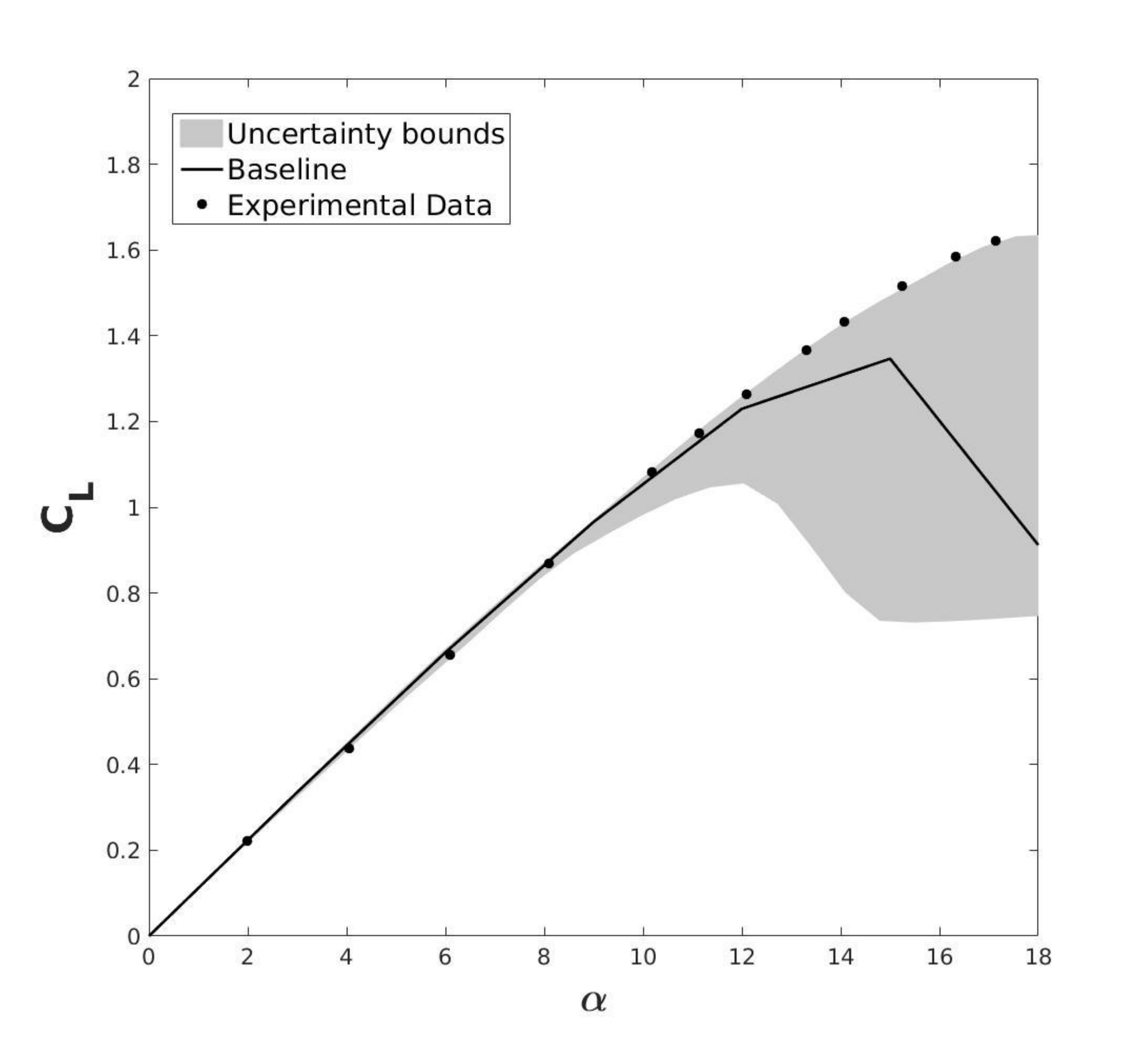}
\caption{Variation in the coefficient of lift $C_L$ with the angle of attack $\alpha$\label{fig:00121}}
\end{figure}

In Fig \ref{fig:00121}, we outline the variation in the coefficient of lift with the angle of attack $\alpha$. At low angles of attack, there is almost no discernible difference between the RANS predictions and the experimental data. Accordingly, here the uncertainty bounds are negligible. At higher angles of attack closer to stall, there is substantial discrepancy between the RANS predictions and the high fidelity data. For these values of $\alpha$, the uncertainty bounds are substantial as well. At all values of $\alpha$, the uncertainty bounds envelope the experimental data. 

\begin{figure}
\center
\subfigure[\label{fig:00122a} Coefficient of pressure variation over the upper surface of the airfoil at $\alpha=10^{\circ}$] %
  {\includegraphics[width=0.8\textwidth]{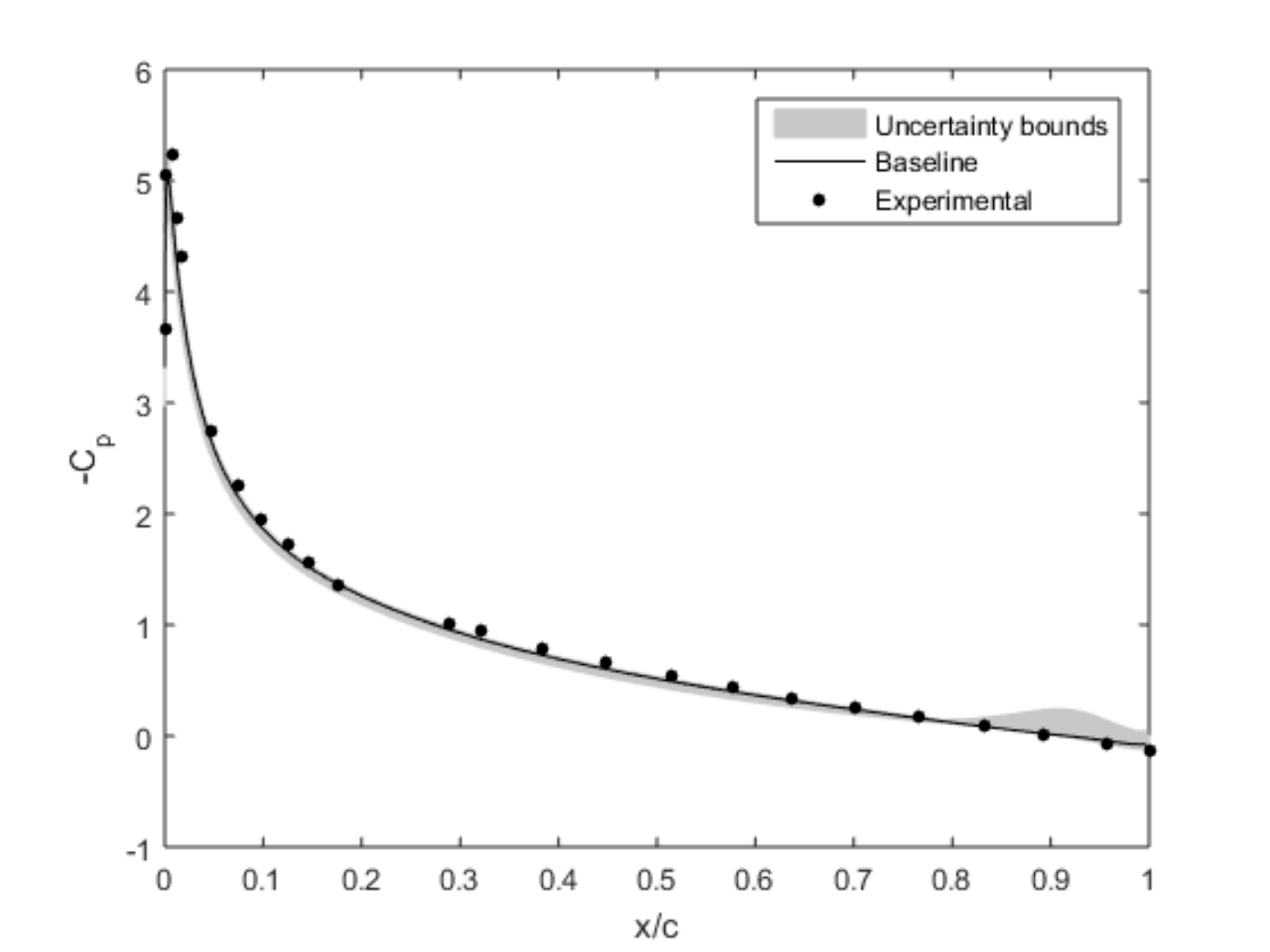}}
\subfigure[\label{fig:00122b}Coefficient of pressure variation over the upper surface of the airfoil at $\alpha=15^{\circ}$] %
  {\includegraphics[width=0.8\textwidth]{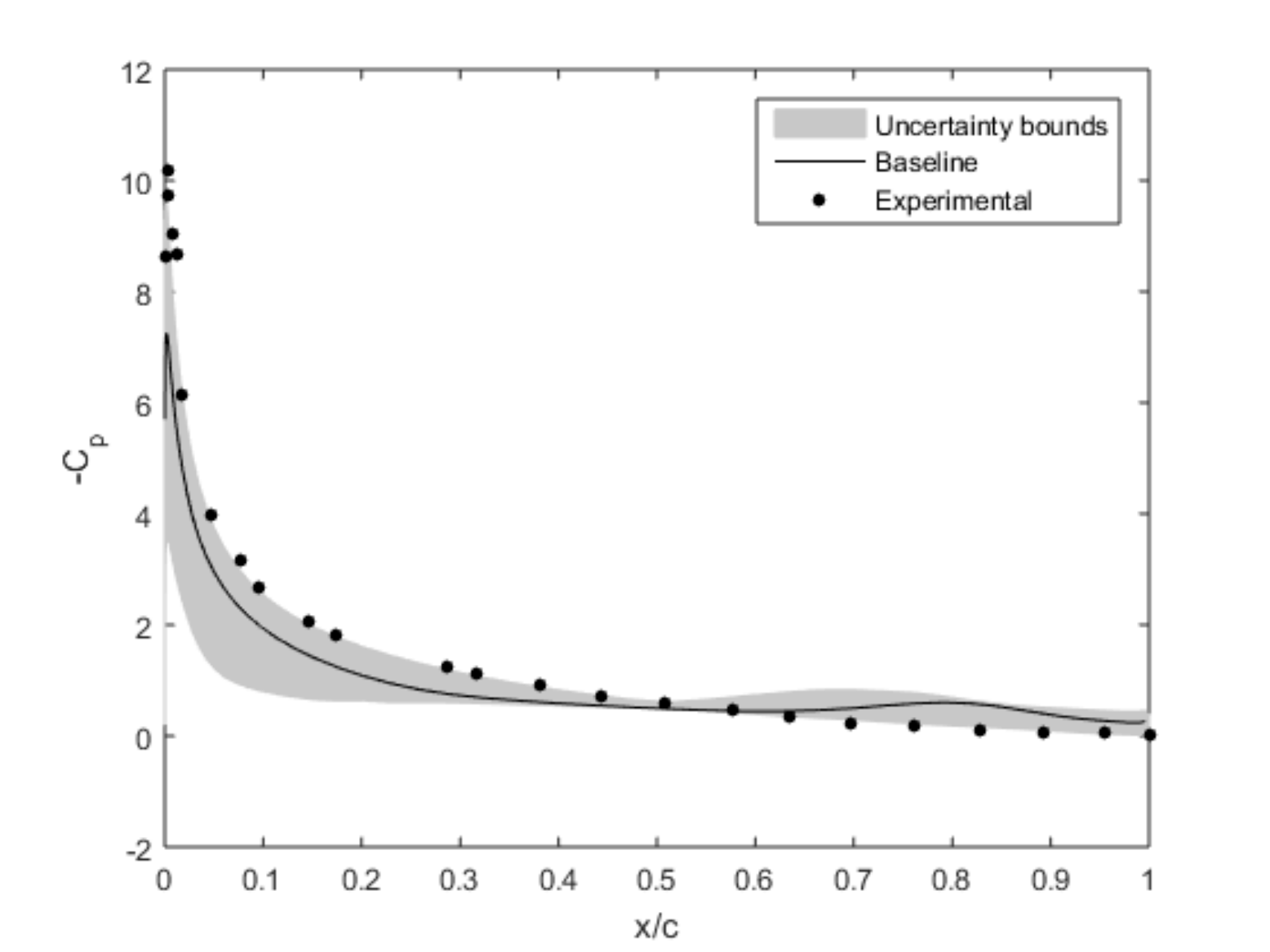}}
\caption{Variation in coefficient of pressure over the airfoil at different angles of attack.}
\end{figure}

Fig. \ref{fig:00122a} and \ref{fig:00122b} outline the variation in the coefficient of pressure ($c_p$) variation along the upper surface of the airfoil at angles of attack $\alpha=10^{\circ}$ and $15^{\circ}$ respectively. Away from the stall at $\alpha=10^{\circ}$, the RANS predictions are in good agreement with the experimental data. Accordingly, the uncertainty bounds are negligibly small. At $\alpha=15^{\circ}$, there is significant discrepancy between the RANS predictions and the experimental data. Here, the uncertainty bounds are sizable and envelope the experimental data. 

\subsection{Turbulent flow over a three-element High-lift airfoil}

In this test case, we investigate the turbulent flow over a McDonnell Douglas Aerospace (MDA) single-flap, three-element airfoil. The flap rigging used corresponds to the 30P/30N designated by MDA. The results correspond to the case with Mach number, $Ma=0.2$; Reynolds number, $Re=5 \times 10^6$ with an angle of attack of $\alpha=8^{\circ}$. We use the experimental data of \cite{chin1993}. 

\begin{figure}
\center
\includegraphics[width=0.8\textwidth]{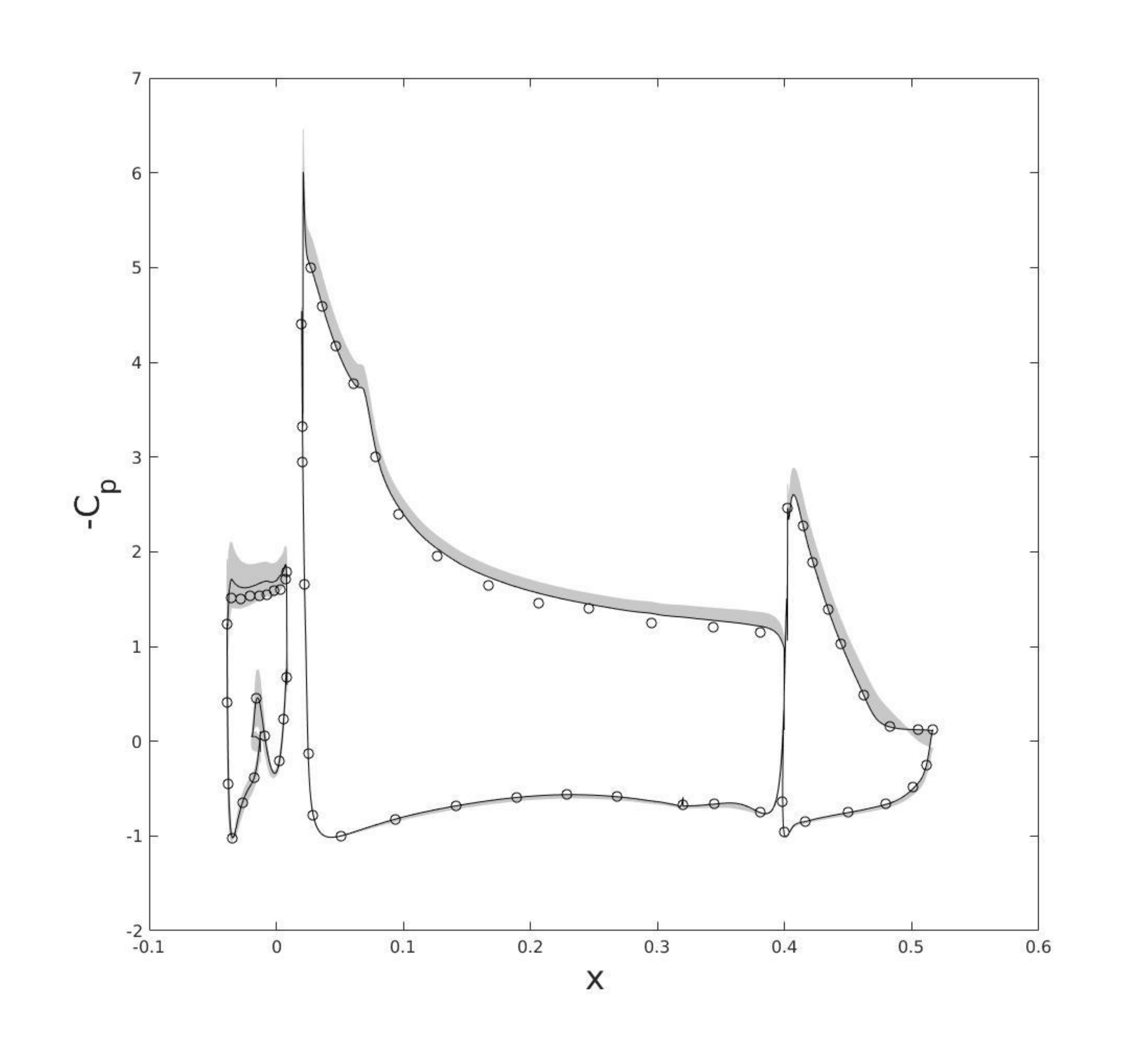}
\caption{Pressure coefficient for the 30P30N configuration\label{fig:30p30n}}
\end{figure}

We outline this case as a test against the false positive. In cases where there is significant discrepancy in the RANS predictions, the uncertainty bounds should exhibit the same. However, in cases where the RANS predictions are accurate, having spurious uncertainty bounds that are significant in their extent would be misleading and would correspond to a false positive. \cite{klausmeyer1997} have tested this flow case for a range of RANS models and have found the RANS predictions to be accurate. In such a scenario, ideally, we would expect the uncertainty bounds to be negligible at most locations along the airfoil sections. Fig. \ref{fig:30p30n} outlines the distribution of the coefficient of pressure ($C_p$) on the surfaces of the different sections. The discrepancy between the RANS predictions and the experimental data is very small over the main element and the flap of the airfoil. Accordingly, the uncertainty bounds are negligibly small over these zones. The upper surface of the slat exhibits an appreciable amount of discrepancy between RANS predictions and the high-fidelity data. The uncertainty bounds over this surface are substantial and are able to envelope the experimental data.

\section{Summary and Conclusions}

The objective of this article is to introduce the EQUiPS (Enabling Quantification of Uncertainty in Physics Simulations) module for the SU2 suite. The EQUiPS module focuses specifically on estimation of uncertainties and errors arising due to turbulence closure models. The theoretical foundations of the module and its computational framework were detailed. The validation results of the module across a set of flow cases were outlined. These flow cases include benchmark flows along with flows pertinent to aerospace design problems. It was found that in all cases, the uncertainty estimates of the module were able to account for a significant portion of the discrepancy between RANS predictions and high fidelity data. The interval estimates engendered were able to envelope most of the high fidelity data. Furthermore, the uncertainty bounds generated exhibit reasonable degrees of both precision and recall. Explicitly, in cases with significant discrepancy between RANS predictions and high fidelity data, the uncertainty bounds are sizable and envelop most of the high fidelity data. However, in cases where there is a high degree of agreement between RANS predictions and high fidelity data, the uncertainty bounds are accordingly negligibly small. This feature increases confidence in the performance of the module and its potential utility as a tool for aerospace design problems. 

In addition to this validation, we have carried out code-to-code comparisons of the SU2 implementation against our propriety code (the results from the latter have been reported earlier in \cite{mishra, mishra2017}) for a range of different flow cases. The module and the ancillary code are is freely available to the community, so that developers may contribute to the code and further improve the features, the functionalities and capabilities of the module.

\bibliographystyle{aiaa}
\bibliography{AIAAEQUiPSRefs}

\end{document}